# The Role of Chain Entropy in an Analytic Model of Protein Binding in Single-DNA Stretching Experiments

Pui-Man Lam[*]
Physics Department, Southern University
Baton Rouge, Louisiana 70813
and
Richard M. Neumann[+]
254 North Maple Street
Florence, Massachusetts 01062

Abstract: We show that the simple analytical model proposed by Zhang and Marko (Phys. Rev. E **77**, 031916 (2008)) to illustrate Maxwell relations for single-DNA experiments can be improved by including the zero-force entropy of a Gaussian chain. The resulting model is in excellent agreement with the discrete persistent-chain model and is in a form convenient for analyzing experimental data.

## I.  Introduction

It is well known that the binding of proteins to DNA may increase (or decrease) the effective persistence length of the DNA. For instance, in the packaging of centimeter-size DNA into microscopic-size chromosomes, the persistence length of DNA is very much reduced. Protein binding on DNA also plays a fundamental role in many cellular and viral functions, including gene expression, physical chromosome organization, chromosome replication, and genetic recombination [1-3]. Therefore understanding the effect of protein binding on DNA is an important topic in cell biology.  Of particular value in such studies [4-9] are micromanipulation techniques, which use magnetic tweezers to measure the extension-versus-force characteristics of single DNA molecules in solution in the presence of known concentrations of proteins.

In such experiments, it is of interest to determine the number of proteins bound to the DNA.  This is a relatively easy task when a protein binds to specific DNA targets of defined sequence and can be accomplished, for instance, by observing the number of jumps in the apparent DNA extension at a fixed force [4, 7, 10] or by detecting fluorescently labeled proteins [7].  However the bound-protein determination becomes more difficult when the protein binding on the DNA is nonspecific, i.e., when binding can occur at any site along a DNA strand.  To address this problem, Zhang and Marko [8] derived analogs of the Maxwell relations of classical thermodynamics which permit one to calculate the change in the number of bound proteins from measurements of DNA extension as a function of bulk protein concentration and applied force. This method has been applied by Liebesny et al. [9] to analyze their extension-versus-force measurements of $\lambda$ DNA in the presence of $\lambda$ repressor protein (CI).

More rigorous theoretical models such as the discrete persistent chain (DPC) [7, 10] exist that describe binding of proteins on DNA. In the DPC model, one considers a discrete realization of the semiflexible DNA polymer with length *L*, made up of a series of segments of length *d*. The orientation of each segment is described by a unit vector and the rigidity of the polymer derives from the interaction of nearest neighbor unit vectors, which favors having the nearest-neighbor unit vectors pointing in the same direction. The interaction strength is proportional to the persistence length which has a value *A* at a site not bound by a protein and a value *A'* at a site bound by a protein. Because the theory considers the protein binding as taking place in solution, a chemical potential ***m*** for the binding of the protein at that site must also be included, where ***m*** > 0 corresponds to an attractive interaction and ***m*** < 0 corresponds to a repulsive interaction of the protein with the DNA. Because the solution of the DPC model is involved, thereby rendering its application to the analysis of experimental data somewhat inconvenient, it would be desirable to show that the simple and more convenient-to-use analytic model proposed by Zhang and Marko is equivalent to the DPC model. Here we show that the two models are indeed equivalent if the entropy of the chain at zero external force is included in the Zhang-Marko (Z-M) model.

In section II we briefly describe the Z-M model and point out its deficiency at small external forces. In section III we describe the zero-force correction to the Z-M model. Section IV is the Conclusion.

**II. A Simple Thermodynamic Model**

Here we briefly describe the Z-M model [8]. Suppose the DNA is of length $L$ in which there are distinct binding sites each of length $d$, for a total of $L/d$ noninteracting sites. The binding of a protein at such a site changes the maximum extension of the binding site from $d$ to $d'$. For a bare DNA strand without any protein, the force-extension curve can be approximated by the formula

$$f(X) = (kT/A)[X/L + (1/4)(1-X/L)^{-2} - 1/4] \qquad (1)$$

where $X$ is the average extension in the direction of the force $f$, $A$ is the persistence length of the DNA and $k$ is Boltzmann's constant. It is convenient to define the Legendre transform as

$$Lg(f) = fX - W(X) \qquad (2)$$

where

$$W(X) = \int_0^X dX' f(X') \qquad (3)$$

is the work done in extending the polymer. Similarly, the function $h(f)$ is defined as

$$h(f) = g(f, A') \qquad (4)$$

where $A'$ is the persistence length when the DNA is bound by proteins.

The partition function for the whole molecule is readily calculated from the two possible states for each binding site:

$$Z = (e^{bgd} + e^{b(m+hd')})^{L/d} \qquad (5)$$

where $m$ is the chemical potential for the binding of a protein at a site.

The average extension is easily calculated:

$$<X> = kT \frac{\partial \log Z}{\partial f} = L \frac{\partial g}{\partial f} + L \frac{\frac{d'}{d} \frac{\partial h}{\partial f} - \frac{\partial g}{\partial f}}{\left[1 + e^{b(gd-m-hd')}\right]} \qquad (6)$$

as is the average number of bound proteins:

$$<N> = kT \frac{\partial \log Z}{\partial m} = \frac{L}{d} \frac{1}{\left[1 + e^{b(gd-m-hd')}\right]} \qquad (7)$$

An unphysical effect of this model can be most easily noted in the limit $f = 0$. In this limit, since the functions $g$ and $h$ are both zero, the average number of proteins as given by Eqn. (7) becomes independent of the persistence lengths $A$ and $A'$. This is in disagreement with the results of the DPC model [7,10].

### III. Zero-force Entropy Correction to the Z-M Model

Physically, at zero force, the end-to-end distance of the polymer should depend on the persistence lengths $A$ and $A'$. The average number of bound proteins should depend on the end-to-end distance and therefore should depend also on $A$ and $A'$, even for $f = 0$. In order to take this into account, we consider the entropy of a Gaussian chain of contour length $L$ and segment length $d$, constrained to have a fixed end-to-end distance $r$. The entropy $S$ of this chain had been calculated by Neumann a long time ago [11]:

$$S = const + 2k \ln r - k \frac{3}{2Ld} r^2 \qquad (8)$$

The entropic force between the ends is given by [11,12]

$$f_s = -T \frac{\partial S}{\partial r} = -T \left[ \frac{2k}{r} - \frac{3k}{Ld} r \right] \qquad (9)$$

Setting the entropic force to zero determines the most probable end-to-end separation $r_p$:

$$r_p = \sqrt{\frac{2Ld}{3}} \qquad (10)$$

Substituting this value into Eqn. (8), one obtains the zero-force entropy of a Gaussian chain

$$S_0 = const + k \ln r_p^2 - k \qquad (11a)$$

We will use the same form of the zero-force entropy for the polymers characterized by persistence lengths $A$ and $A'$, with also

$$S_0' = const + k \ln r'^2_p - k \qquad (11b)$$

where $r_p, r'_p$ are the most probable end-to-end distances of the bare and protein-bound chains, respectively.

In the Z-M model, according to Eqn. (5) the partition function of the whole chain is given by the partition function of a single segment raised to the power of the total number of segments in the chain. In the same spirit, in order to calculate the zero-force entropy correction to the Z-M model, we have to determine the entropy of a segment of chain with contour length $d$ or $d'$. Starting with the relationship $r_p^2 = 2AL$, which holds for a worm-like chain [13] of contour length $L$, the corresponding relationships for the segments of contour length $d$ and $d'$ are, respectively,

$$r_s^2 = 2Ad = 2l_0 d^2, \quad r'^2_s = 2A'd' = 2l_1 d'^2 \qquad (12)$$

where $r_s^2$ and $r_s'^2$ are the squares of the most-probable end separations for the respective chain segments, and $l_0$ and $l_1$ are dimensionless ratios defined according to

$$A = l_0 d, \quad A' = l_1 d'. \qquad (13)$$

Using Eqns. (11a) and (11b) with $r_s^2$ and $r_s'^2$ replacing $r_p^2$ and $r'^2_p$, respectively, the corrections to the partition function due to the zero-force entropies of the bare and protein-bound chain segments are

$$\exp(-TS_0/kT) \sim \exp(-\ln(l_0 d^2)) \sim (l_0 d^2)^{-1} \qquad (14a)$$

$$\exp(-TS'_0/kT) \sim \exp(-\ln(l_1 d'^2)) \sim (l_1 d'^2)^{-1} \qquad (14b)$$

The partition function is now given by

$$Z \sim \left[(l_0 d^2)^{-1} e^{bgd} + (l_1 d'^2)^{-1} e^{b(m-hd')}\right]^{\frac{L}{d}} \qquad (15)$$

The average extension is given by

$$<X> = k_B T \frac{\partial \log Z}{\partial f} = L \frac{\partial g}{\partial f} + L \frac{\frac{d'}{d} \frac{\partial h}{\partial f} - \frac{\partial g}{\partial f}}{\left[1 + \frac{l_1 d'^2}{l_0 d^2} e^{b(gd - m - hd')}\right]} \qquad (16)$$

For $m = 0, f = 0$, this reduces to

$$<X> = L \frac{l_1 d'^2}{l_0 d^2 + l_1 d'^2} \frac{\partial g}{\partial f} + L \frac{d'}{d} \frac{l_0 d^2}{l_0 d^2 + l_1 d'^2} \frac{\partial h}{\partial f}. \qquad (17)$$

The average number of proteins is given by

$$<N> = k_B T \frac{\partial \log Z}{\partial m} = \frac{L}{d} \frac{1}{\left[1 + \frac{l_1 d'^2}{l_0 d^2} e^{b(gd - m - hd')}\right]} \qquad (18)$$

For $m = 0, f = 0$, this reduces to

$$<N> = \frac{L}{d} \frac{l_0 d^2}{l_0 d^2 + l_1 d'^2} \qquad (19)$$

The cross derivative, i.e., Maxwell relation, is given by

$$kT \frac{\partial <N/L>}{\partial f} = kT \frac{\partial <X/L>}{\partial m} = \frac{\left[\frac{d'}{d} \frac{\partial h}{\partial f} - \frac{\partial g}{\partial f}\right] \frac{l_1 d'^2}{l_0 d^2} e^{b(gd - m - hd')}}{\left[1 + \frac{l_1 d'^2}{l_0 d^2} e^{b(gd - m - hd')}\right]^2} \qquad (20)$$

This shows that the revised model continues to observe the Maxwell relation for $N \, dm + X \, df$. Using these equations we have calculated the force-extension curves and the fraction (m) of bound proteins for the case $d = d'$. Figures 1a and 1b show these quantities for the case $l_0 = 50$, $l_1 = 25$. The solid lines were obtained using the corrected Z-M model, the dashed lines using the uncorrected Z-M model, and the crosses using the DPC model [10]. In figures 2a and 2b the same quantities are shown for the case $l_0 = 50$, $l_1 = 100$. It is clear that the findings from the corrected Z-M model are in excellent agreement with those from the DPC model, while differing considerably from the findings of the uncorrected Z-M model.

## IV. Conclusion

We have shown that by taking into account the entropy at zero force, originally derived for an ideal Gaussian chain, the Z-M model can be brought into accord with the DPC model. Because the corrected Z-M model incorporates the most desirable features from both the DPC and Z-M models, we believe that it should be the model of choice when analyzing experimental data. It combines the rigor of the DPC model with the convenience of using the Z-M model, which is formulated in terms of analytic functions. Furthermore the continuous nature of the corrected Z-M model makes it more suitable, compared with the DPC model, when describing the effects of twist and torque.

Finally, it is worth emphasizing that the current model is only appropriate for describing proteins which effect a change in local persistence length and segment length when binding in a non-sequence-specific manner along the double helix. The model does not describe proteins which kink or otherwise deform the double helix.

* puiman_lam@subr.edu
+ richard.neumann@post.harvard.edu

**Acknowledgement.** Research supported by the Louisiana Board of Regents Support Fund Contract Number LEQSF(2007-10)-RD-A-29 and Louisiana Epscor EPS-1003897 and NSF(2010-15)-RII-SUBR.

Figure (a): Plot of $\langle X \rangle / L$ (dimensionless units) versus $fd/kT$ (dimensionless units) on a semi-log scale. Curves are labeled $\beta\mu = -4$, $\beta\mu = -2$, $\beta\mu = 0$, $\beta\mu = 2$, $\beta\mu = 4$. Parameters: $l_0 = 50$, $l_1 = 25$, $d = d'$.

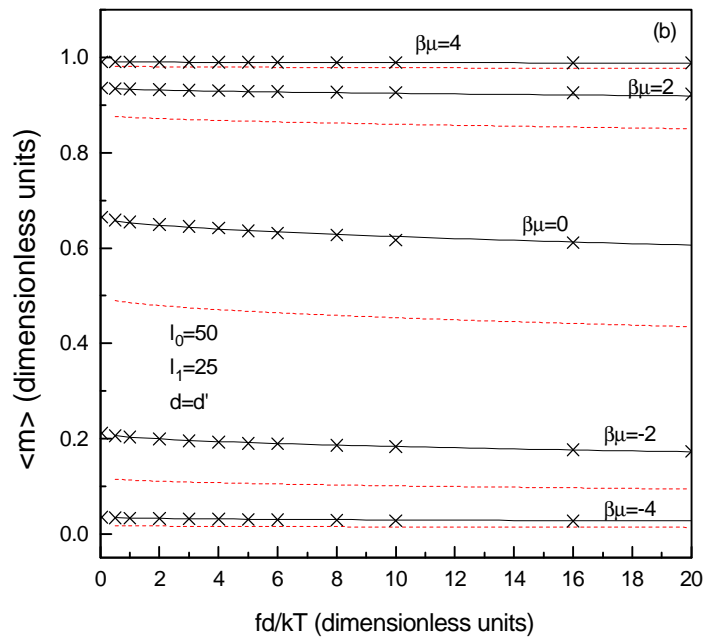

Fig. 1: (Color online) (a) Force-extension curves and (b) fraction of bound proteins as a function of force for $l_0 = 50$, $l_1 = 25$, $d = d'$. Solid lines are corrected Z-M model results, dashed lines (in red) are for uncorrected Z-M model results, and crosses are results of DPC model.

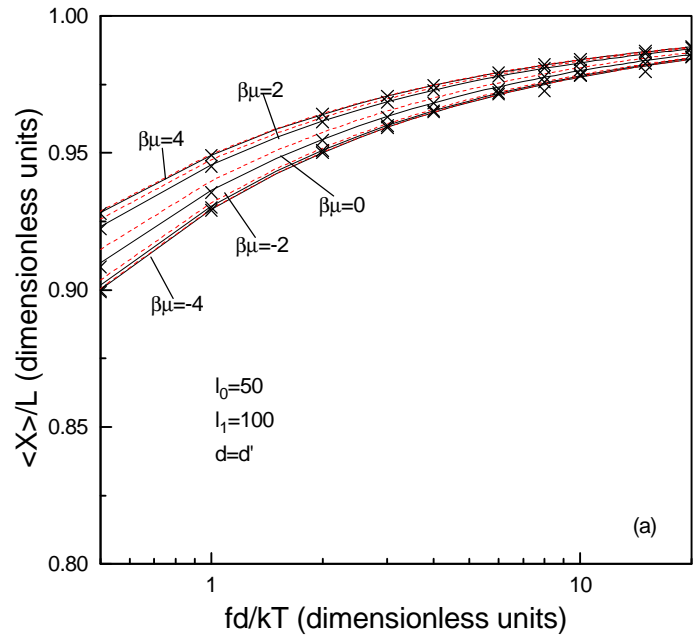

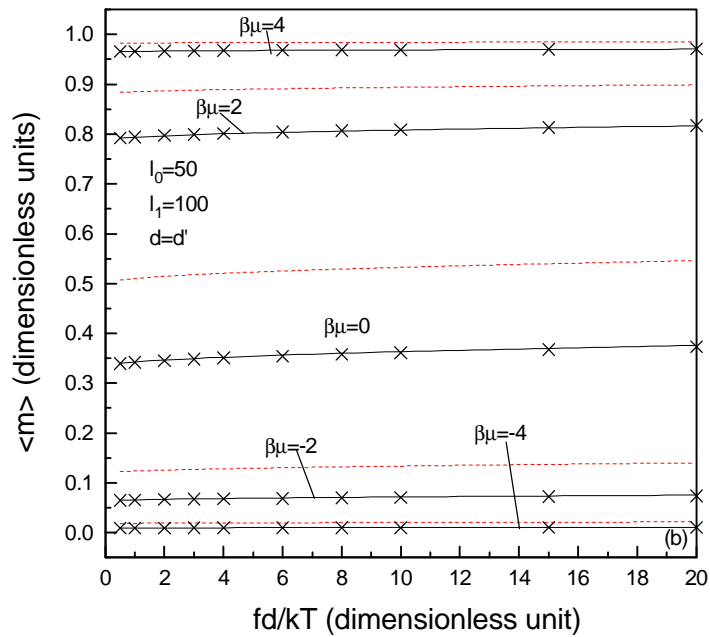

Fig. 2: (Color online) (a) Force-extension curves and (b) fraction of bound proteins as function of force, for $l_0 = 50$, $l_1 = 100$, $d = d'$. Solid lines are corrected Z-M model results, dashed lines (in red) are for uncorrected Z-M model results, and crosses are results of DPC model.